\documentclass[twocolumn,prb,10pt,aps,longbibliography,superscriptaddress]{revtex4-2}

\usepackage{graphicx}
\usepackage{grffile}
\usepackage{amsmath}
\usepackage{amssymb}
\usepackage{amsfonts}
\usepackage{dsfont}
\usepackage{dcolumn}
\usepackage{bbm}
\usepackage{bm}
\usepackage{mathtools}
\usepackage{braket}
\usepackage{placeins}
\usepackage{physics}
\usepackage{siunitx}
\usepackage[unicode]{hyperref}
\usepackage[nameinlink]{cleveref}
\Crefname{section}{Sec.}{Secs.}
\Crefname{equation}{Eq.}{Eqs.}
\Crefname{figure}{Fig.}{Figs.}
\Crefname{tabular}{Tab.}{Tabs.}

\usepackage{bbold} 
\usepackage{nicefrac}
\usepackage{tikz}
\usetikzlibrary{arrows.meta}
\usepackage{lipsum}  
\usepackage{xcolor}

\usepackage[makeroom]{cancel}

\definecolor{nat_green}{HTML}{43B02A}

\definecolor{tdo_green}{HTML}{83B818}
\definecolor{tdo_darkgreen}{HTML}{839A00}

\definecolor{rw_red}{HTML}{C8102E}
\definecolor{rw_darkred}{HTML}{971B2F}

\definecolor{med_blue}{HTML}{00A3E0}
\definecolor{med_darkblue}{HTML}{0061A0}

\definecolor{tdo_orange}{HTML}{D98207}
\definecolor{tdo_darkorange}{HTML}{CA7406}

\renewcommand\vec{\mathbf}

\usepackage{xstring}
\newcommand{\refcite}[1]{%
\begingroup
\def\tempx{0}%
  \StrCount{#1}{,}[\tempx]%
  \ifnum\tempx > 0 
  Refs.~%
  \else
  Ref.~%
  \fi
\endgroup
\cite{#1}%
}

\newcommand{\bes}{\begin{subequations}}
\newcommand{\ees}{\end{subequations}}
\newcommand{\be}{\begin{equation}}
\newcommand{\ee}{\end{equation}}

\newcommand{\im}{\mathrm{i}}

\NewDocumentCommand\alp{ s g }{ \ensuremath{ \IfBooleanTF#1
{ \alpha^{\dagger}_{#2}    } { \alpha^{\phantom{\dagger}}_{#2} } } }

\NewDocumentCommand\bet{ s g }{ \ensuremath{ \IfBooleanTF#1
{ \beta^{\dagger}_{#2}    }{ \beta^{\phantom{\dagger}}_{#2} } }} 

\NewDocumentCommand\alpeff{ s g }{ \ensuremath{ \IfBooleanTF#1
{ \alpha^{\dagger}_{#2}    } { \alpha^{\phantom{\dagger}}_{#2} } } }

\NewDocumentCommand\beteff{ s g }{ \ensuremath{ \IfBooleanTF#1
{ \tilde\beta^{\dagger}_{#2}    }{ \tilde\beta^{\phantom{\dagger}}_{#2} } }}

\NewDocumentCommand\abos{ s g }{ \ensuremath{ \IfBooleanTF#1
{ a^{\dagger}_{#2}    } { a^{\phantom{\dagger}}_{#2} } } }

\NewDocumentCommand\bbos{ s g }{ \ensuremath{ \IfBooleanTF#1
{ b^{\dagger}_{#2}    } { b^{\phantom{\dagger}}_{#2} } } }

\DeclareMathOperator{\sign}{sign}

\NewDocumentCommand\vect{g} {\ensuremath{  \bm{#1}  } }
\NewDocumentCommand\mat{g} {\ensuremath{  \text{\textbf{#1}}  } }

\NewDocumentCommand\cexpval{g}{\ensuremath{  \Braket{#1}}_0 }
\NewDocumentCommand\norord{g}{\ensuremath{ :{#1}: } }

\newcommand{\mbzO}{\ensuremath{  N_0 } }
\newcommand{\mbz}{\ensuremath{ N_{\emptyset}}}

\newcommand{\gs}{\ensuremath{E}}
\newcommand{\gsPerSite}{\ensuremath{e}}

\newcommand{\gap}{\ensuremath{\Delta}}
\newcommand{\disp}{\ensuremath{\omega}}

\newcommand{\boundstates}{\ensuremath{ \tau_{i}}}

\NewDocumentCommand\boundstate{g}{\ensuremath{ \tau(#1) } }

\begin{document}
    \title{Continuous similarity transformation for critical phenomena: easy-axis antiferromagnetic XXZ model}

    \author{Matthias R.\ Walther}
    \email{matthias.walther@fau.de}
    \affiliation{Department of Physics, Friedrich-Alexander-Universit\"{a}t Erlangen-N\"urnberg, Staudtstra\ss{}e 7, 91058 Erlangen, Germany}

    \author{Dag-Bj\"orn Hering}
    \email{dag.hering@tu-dortmund.de}
    \affiliation{Condensed Matter Theory, 
    Technische Universit\"{a}t Dortmund, Otto-Hahn-Stra\ss{}e 4, 44221 Dortmund, Germany}

    \author{G\"otz S.\ Uhrig}
    \email{goetz.uhrig@tu-dortmund.de}
    \affiliation{Condensed Matter Theory, 
    Technische Universit\"{a}t Dortmund, Otto-Hahn-Stra\ss{}e 4, 44221 Dortmund, Germany}

    \author{Kai P.\ Schmidt }
    \email{kai.phillip.schmidt@fau.de}
    \affiliation{Department of Physics, Friedrich-Alexander-Universit\"{a}t Erlangen-N\"urnberg, Staudtstra\ss{}e 7, 91058 Erlangen, Germany}

    \date{\textrm{\today}}

    \begin{abstract}
        We apply continuous similarity transformations (CSTs) to the easy-axis antiferromagnetic XXZ-model on the square lattice. 
        The CST flow equations are truncated in momentum space by the scaling dimension $d$ so that all contributions with $d\le 2$ are taken into account.
        The resulting quartic magnon-conserving effective Hamiltonian is analyzed in the zero-, one-, and two-magnon sector.
        In this way, a quantitative description of the ground-state energy, the one-magnon dispersion and its gap as well as of two-magnon bound states is gained for anisotropies ranging from the gapped Ising model to the gapless Heisenberg model.
        We discuss the critical properties of the gap closing as well as the evolution of the one-magnon roton mininum. 
        The excitation energies of two-magnon bound states are calculated and their decay into the two-magnon continuum is determined via the inverse participation ratio.
    \end{abstract}

    \maketitle

    \section{Introduction}
    \label{s:introduction}
        The collective behavior of interacting quantum matter is an important topic in condensed matter physics as well as quantum optics over the last decades, since more and more intriguing facets of quantum systems are discovered giving rise to unexpected properties of quantum materials and fascinating perspectives in quantum technological applications.
        Such quantum behavior manifests itself most prominently at low temperatures where quantum fluctuations dominate and therefore zero-temperature quantum phases and 	phase transition with emergent properties are a relevant subject of current research 	\cite{sachd99}. 
        Here two-dimensional systems are most challenging, since, unlike in one dimension, no analytical nor generic numerical solutions are available and, 	unlike in three dimensions, quantum fluctuations are still important. 
        At the same time two-dimensional correlated quantum systems are known for exciting physical properties such as high-$T_{\rm c}$ superconductivity, the fractional quantum Hall effect as well as highly entangled topological order.      
       
        One paradigmatic microscopic model is the antiferromagnetic spin-$\tfrac{1}{2}$ square lattice Heisenberg model, which has been intensively studied, in particular since it represents the relevant low-energy description of the quantum magnetism in the undoped cuprate superconductors \cite{manou91}.
        While properties of the long-range ordered N\'eel state and the gapless low-energy magnon excitations are well known \cite{Auerbach_1994}, a quantitative understanding of the high-energy part of the one-magnon dispersion displaying a characteristic roton minimum as well as of dynamical 	correlation functions including important contributions of magnon continua have only been achieved in recent years \cite{Powalski_2015,Powalski_2018,shao17,Verresen_2018,ferra18}. In parallel, the latter quantities become more and more important experimentally due to the continuously increasing resolution of various spectroscopic techniques such as inelastic neutron scattering, terahertz spectroscopy, or resonant inelastic x-ray scattering. 
   
        A quantitative description of the magnon excitation spectrum in the  Heisenberg model on the square lattice requires a proper treatment of the magnon-magnon interaction.
        Technically, this can be achieved by continuous similarity transformations (CSTs) using the scaling dimension of operators as truncation criterion of the associated flow equations \cite{Powalski_2015,Powalski_2018}.
        The CST yields an effective Hamiltonian in momentum space which is block-diagonal in the dressed magnon quasi-particles and which allows one to quantitatively access one- and multi-magnon excitation energies.
        It is a natural next step to ask whether CSTs can be applied successfully when modifying the model away from the gapless Heisenberg point.
        The easy-axis antiferromagnetic XXZ model on the square lattice represents one such modification which tunes continuously from the isotropic Heisenberg model to the classical Ising model with local and gapped magnon excitations.
        The magnon excitation spectrum of the XXZ model has been investigated by several means, in particular with high-order series expansions (SE) \cite{Zheng_1991,Zheng_2005,Dusuel_2010}, diagrammatic spin wave theory \cite{Hamer_1992,Hamer_2009}, density matrix renormalization group (DMRG) \cite{Verresen_2018}, and quantum Monte Carlo simulations (QMC) \cite{Sandvik_1997,Sandvik_2001}.
        Apart from the gap closing transition, a notable feature is the appearance of two-magnon bound states in large parts of the phase diagram, which are not present at the Heisenberg point where there is only a resonance of the Higgs-like amplitude mode within the magnon continuum \cite{weidi15,Powalski_2018}.
        Interestingly, in coupled XXZ spin ladder compounds the finite easy-axis spin anisotropy has lead to the experimental discovery of the Higgs amplitude mode near a quantum critical point \cite{hong17,Ying_2019}.       

        This paper is structured as follows.
        In \Cref{s:model} we describe the relevant properties of the easy-axis XXZ model on the square lattice.
        A description of the CST as well as of technical aspects are provided in \Cref{s:methods}; this section can be skipped by those who want to focus on the results. Results for the ground-state energy, the one-magnon dispersion, and the two-magnon sector are presented and discussed in \Cref{s:results}.We conclude our study in \Cref{s:summary}.

    \section{Model}
    \label{s:model}
        We study the antiferromagnetic easy-axis XXZ model on the square lattice as prototypical extension of the isotropic Heisenberg model. 
        Here, we set the lattice constant $a$ to unity.
        The anisotropy induces a finite energy gap which vanishes in the isotropic limit where the  Goldstone theorem applies \cite{Auerbach_1994}.
        Hence, we can study how the gap closes upon approaching the isotropic point and whether binding effects occur.
        To establish a deeper understanding of the performance and the applicability of continuous basis transformations \cite{Powalski_2015,Powalski_2018} for vanishing energy gaps we investigate the XXZ model with easy-axis anisotropies from the Ising limit to the isotropic Heisenberg limit. 
        The Hamiltonian reads
        \begin{align}
            \label{eq:hamiltonian}
            \mathcal{H} &= J \sum_{\Braket{i,j}} \left[S_{i}^{z} S_{j}^z + \lambda \left( S_i^x S_j^x + S_i^y S_j^y \right)\right]
        \end{align}
        where $J > 0$ is the exchange coupling between spins on site ${i}$, $\Braket{i,j}$ represents the sum over pairs of nearest neighbors on the square lattice where each bond is counted  once. 
        The spin anisotropy  is introduced by $\lambda \in [0,1]$ with the $z$-axis being the easy axis.
        For $\lambda=0$ the XXZ model reduces to the Ising model which is exactly solvable \cite{Onsager_1944}. 
        It displays a long-range ordered ground-state, a gapped, completely flat dispersion and a two-particle sector with four nearest-neighbor bound states per site in one sublattice.
        The Ising model is a frequently used starting point for perturbative treatments of Heisenberg and XXZ models, see for instance Ref.\ \cite{Zheng_2005, Oitmaa_2006, Dusuel_2010}.
        This Ising phase continues adiabatically, i.e.,  without a quantum phase transition, up to $\lambda =1$. 
        At this point the symmetry of \Cref{eq:hamiltonian} changes from $\mathds{Z}_2\times\text{U}(1)$  to $\text{SU}(2)$ of all spin rotations.
        Here, at $\lambda=1$, the XXZ model becomes the isotropic Heisenberg model still displaying long-range magnetic order which breaks its \emph{continuous} symmetry spontaneously so that the Goldstone theorem implies the existence of gapless elementary excitations, the magnons \cite{Auerbach_1994}.
        Due to the absence of an energy gap, multi-magnon continua exist starting right at the energies of single magnons extending to higher energies.
        No bound states are known at the isotropic point, but the attractive interaction between magnons leads to considerable shift of spectral weight to lower energies \cite{Powalski_2015,Powalski_2018}.
        We focus on the following physical quantities:

        \paragraph{Ground-state energy} 
        The ground-state in the Ising limit {($\lambda=0$)} is precisely the N\'eel state with a ground-state energy per site $e= -J/2$ that spontaneously breaks the discrete $\mathds{Z}_2$ symmetry. 
        The other limiting case is $\lambda=1$ where the ground-state is the one of the isotropic Heisenberg model.
        In the thermodynamics limit, it spontaneously breaks the SU(2) symmetry.
        The ground-state energy per site obtained from QMC takes the value \mbox{$e^{\text{QMC}}=-0.669437(5)J$} \cite{Sandvik_1997}. 
        This value serves as a first testbed for our approach.
        \paragraph{Dispersion $\disp(\vect{k})$ and single-particle gap $\Delta$}
        For $\lambda = 0$ the dispersion is completely flat $\disp(\vect{k})=2J$ because the elementary excitations, spin flips, are completely immobile.
        For $\lambda > 0$ more and more features emerge.
        In the magnetic Brillouin zone (MBZ), the dispersion displays a minimum at $\vect{k} = (0,0)$ which defines the spin gap $\Delta\coloneqq\disp(\vect{k}=\vect{0})$.
        It closes at the isotropic Heisenberg point $\lambda = 1$ according to Goldstone's theorem.
        The closure of the spin gap follows a square root power law in spin wave theory \cite{Zheng_2005} although series expansions would be consistent for slightly higher exponents as well \cite{Zheng_1991}.
        Additionally, the dispersion along the line from $\vect{k} = (\pi,0)$ to $\vect{k} = (\frac{\pi}{2},\frac{\pi}{2})$ in the Brillouin zone develops a distinct local minimum  at $\vect{k} = (\pi,0)$, called roton minimum, and the global maximum at $\vect{k} = (\frac{\pi}{2},\frac{\pi}{2})$. 
        The roton minimum occurs only in third order spin wave theory \cite{Syromyatnikov_2010}.
        It is understood to be due to magnon-magnon interactions \cite{Powalski_2015,Powalski_2018} and a peculiar cancellations for motion along the diagonals \cite{Verresen_2018}.
        We will analyse the spin gap $\Delta$, the roton minimum, and dispersion maximum in dependence on the anisotropy $\lambda\in [0, 1]$ with a special focus on the critical behavior of $\Delta$ for $\lambda\to1$.

        \paragraph{Two-particle sector}
        The eigenstates which are built from two elementary excitations, i.e., two magnons, are particularly interesting because they reflect the degree of magnon-magnon interaction. 
        For strong interaction relative to the kinetic energy of the magnons, bound states occur in the gapped phase.
        The two-particle spectrum of the Ising model features degenerate bound states at $3J$ stemming from adjacent pairs of spin flips.
        All other states of two magnons have energy $4J$ and are thus highly degenerate.
        For finite $\lambda$, these states evolve into energy continua of scattering states.
        The four bound states exist also for finite $\lambda$, but merge with the continuum for values $\lambda$ close to but smaller than one \cite{Oguchi_1973, Hamer_2009,Dusuel_2010} consistent with quantum Monte Carlo simulations of coupled XXZ spin ladders \cite{Ying_2019}.
        We track these bound states for all $\lambda$ and determine at which values of $\lambda$ the bound states vanish in the two-magnon continuum by analyzing the locality of bound states with the help of the inverse participation ratio (IPR) \cite{Kramer_1993,Malki_2019}.
    \section{Methods and Techniques} 
    \label{s:methods}
        Here we recall the fundamental idea of continuous basis transformations in general and the continuous similarity transformation (CST) in particular which we employ here.
        Further methodological aspects are also recapitulated such as the self-consistent mean-field approach to obtain a good starting point for the CST and the IPR as a measure of locality.
        Technical aspects such as the discretization of the magnetic Brillouin zone (MBZ) and various extrapolations are addressed as well.


        \subsection{Continuous Similarity Transformation (CST)}
				\label{ss:cst}
        We use the CST method in momentum space as introduced in \refcite{Powalski_2015, Powalski_2018}. 
        As in the flow equation approach \cite{Wegner_1994,Knetter_2000,Kehrein_2006}, we introduce an auxiliary flow variable $\ell$ that parametrizes a continuous similarity transformation from $\mathcal{H}(\ell=0) = \mathcal{H}_0$ to an effective Hamiltonian $\mathcal{H}(\ell=\infty) = \mathcal{H}_{\text{eff}}$.
        The flow is governed by the flow equation $\partial_{\ell} \mathcal{H}(\ell) = \left[ \eta(\ell), \mathcal{H}(l) \right]$, where the generator $\eta(\ell)$ determines the course of the flow and its end point, the fixed point.
        We use the quasi-particle conserving generator $\eta_{ij} = (q_{ii}- q_{jj})h_{ij}$ \cite{Knetter_2000, Mielke_1994, Fischer_2010}, where $q_{ii}$ are the eigen values of the operator $\hat{Q}$ counting the number of quasi-particles, here magnons, present in the system.
        Then the final effective Hamiltonian $\mathcal{H}_{\text{eff}}$ has a block diagonal form, where the subspace on which each block acts has a given number of quasi-particles.
        Subspaces with different numbers of magnons do not influence each other; they are disentangled.
        The coefficient in the zero-magnon block is the ground-state energy, the one-magnon block contains the dispersion, and the two-magnon energies are obtained by diagonalizing the two-magnon block.

        Computing $\left[ \eta(\ell), \mathcal{H}(\ell) \right]$ for the flow equation generally yields an infinite number of terms which cannot be handled numerically. 
        Therefore, one has to truncate the flow equations in a systematic and controlled way.
        We follow here the idea to use the scaling dimension as truncation criterion.
        The higher this dimension the less important are the corresponding terms for the physics at low energies.
        For the sine-Gordon model, the operator product expansion implements this idea and allows one to systematically truncate the flow according to Wegner's generator \cite{Kehrein_1999, Kehrein_2001}.
        Recently, the truncation according to scaling dimension justified to truncate the flow to the bilinear and quadrilinear terms in a bosonic Dyson-Maleev re\-pre\-sen\-tation \cite{Dyson_1956,Maleev_1958} of the spins \cite{Powalski_2015, Powalski_2018}.
        At the isotropic point $\lambda=1$, the dispersion encoded in the bilinear terms has dimension 1 while the quadrilinear terms have dimension 2 for essentially constant prefactors. 
        The neglected hexalinear terms would have dimension 4.
        Here, we focus on the gapped spectrum of the anisotropic easy-axis so that the dimension of the bilinear dispersion is 0.
        Hence the same truncation remains justified and we do not change it because the limit $\lambda\to 1$ shall be captured as well. 
  	    Eventually, the flow equations are solved numerically for a discretized MBZ. 
        Details of the numerical methods are presented in \Cref{ss:flow-rod,ss:discrete,ss:extrapol-linear}.
        We will see that the results for the dispersion and the energy gap in \Cref{ss:dispersion} strongly corroborate our approach.
        In addition, we discern and discuss deviations from exact series results \cite{Hamer_2009,Dusuel_2010} for the energies of the bound states, see \Cref{ss:boundstates}.

        \subsection{Self-consistent mean-field approach}
        \label{ss:mf}
		Starting truncated transformations from spin representations in real space has turned out to be difficult and prone to large truncation errors \cite{Schmidt_2006}.
	    Hence, we use the self-consistent mean-field solution in momentum space as initial Hamiltonian even though a variation of this solution can sometimes perform better \cite{Uhrig_2013}.
        The crucial advantage of the self-consistent mean-field approach is that it captures the Goldstone bosons of the isotropic Heisenberg model.
	    To be self-contained we sketch the necessary analytical steps applied to Hamiltonian  \eqref{eq:hamiltonian}.
        A detailed derivation with intermediate results is given in \Cref{a:s:mean_field}.

\begin{figure}[!ht]
            \centering
                \resizebox{0.2\textwidth}{!}{\includegraphics{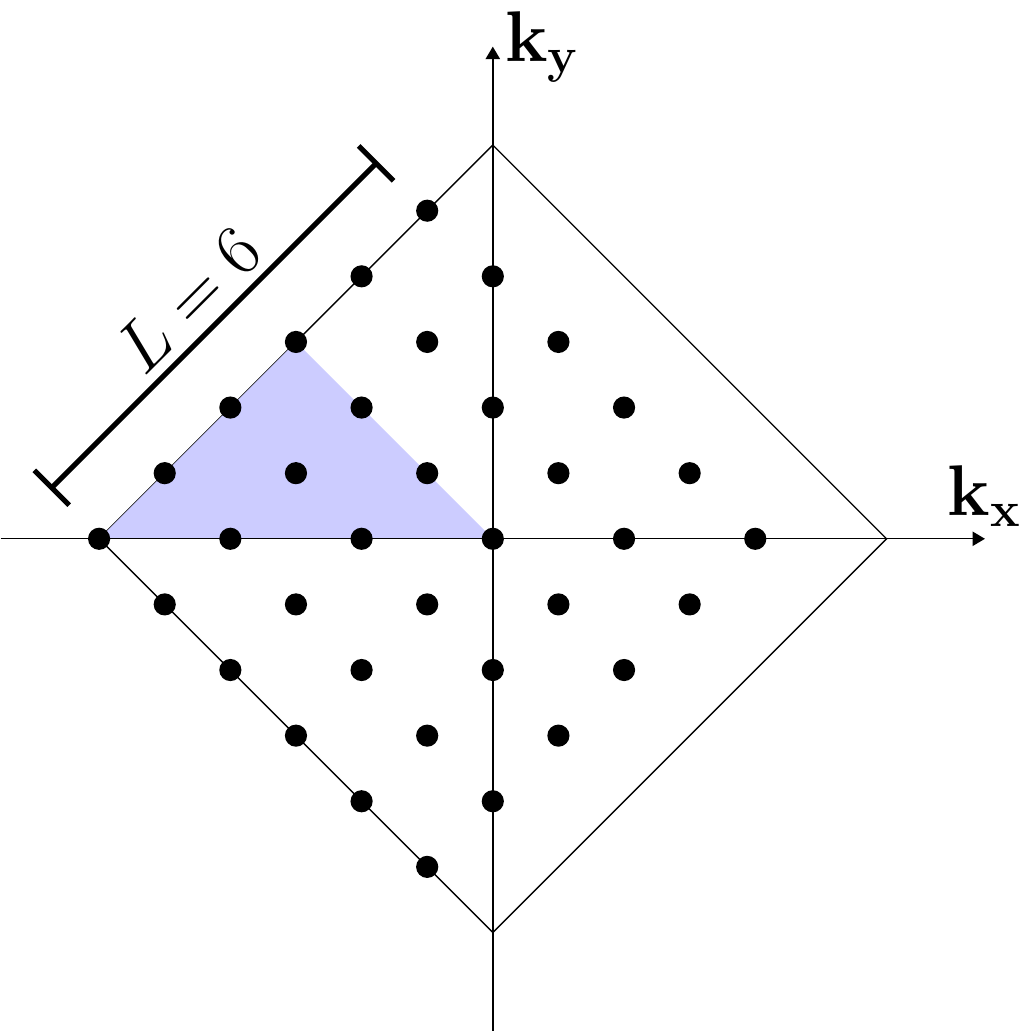}}
                \resizebox{0.2\textwidth}{!}{\includegraphics{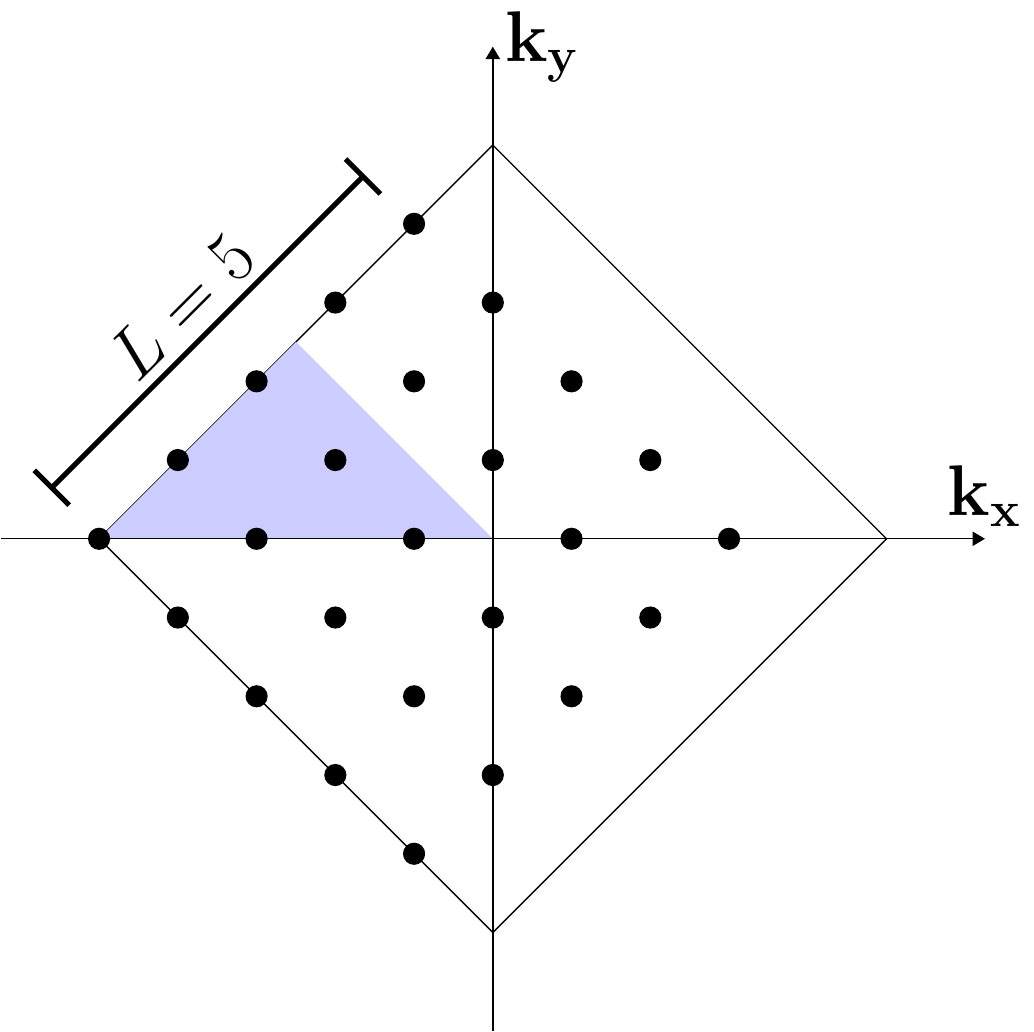}}

                \caption{Magnetic Brillouin zone (MBZ) with different kinds of discretizing mesh. 
            In panel (a) important points such as $\vect{k}=(0,0)$ are sampled directly; this kind of mesh is denoted by $\mbzO$. 
            For $\mbzO$, the calculations for $\lambda < 1$ or the two-magnon sector do not require additional interpolations.
            But at $\vect{k}=(0,0)$ divergences in the flow occur for $\lambda \lesssim 1$. 
            They can be avoided by a mesh of the type shown in panel (b) for odd $L$; this mesh is denoted by $\mbz$.
	        We use it especially for the analysis of the critical behavior of the spin gap.
            Due to the point group symmetries only lattice sites in the marked blue area need to be taken into account.} 
            \label{fig:mbz_mesh}
        \end{figure}

        The ground-state for $0 \le \lambda \le 1$ displays long-range N\'eel order.
        Thus, we choose the same classical N\'eel state with alternating spin-up and spin-down on sublattices A and B, respectively, as the reference state for all $\lambda$.
        We introduce bosonic degrees freedom that represent deviations from this state by the non-Hermitian Dyson-Maleev representation \cite{Dyson_1956,Maleev_1958}.
        Then we perform a standard mean-field decoupling \cite{Takahashi_1989} based on Wick's theorem in real space.
        A Fourier transformation and a subsequent Bogoliubov transformation to the bosonic operators $\alpha^{(\dagger)}_{\vect{k}}$ and $\beta^{(\dagger)}_{\vect{k}}$ diagonalize the Hamiltonian on the bilinear level where the $\alpha$ bosons	refer to the A sublattice and the $\beta$ bosons to the B sublattice.
        The mean-field Hamiltonian reads
        \begin{align}
            \label{eq:startHamil}
            \mathcal{H} = E_0 + \sum_{\vect{k}} \disp_0(\vect{k}) 
	    			\big(\norord{ \alp*{\vect{k}}\alp{\vect{k}}} + 
	    			\norord{\bet*{\vect{k}}\bet{\vect{k}}}\big) + \Gamma_0 + \mathcal{V}_0.
        \end{align}
        The wave vectors $\vect{k}$ are taken from the MBZ depicted in \Cref{fig:mbz_mesh}. The colons $\norord{...}$ indicate normal ordered operators with respect to the mean-field ground-state.
        The exact expressions of the ground-state energy $E_0$, the dispersion $\omega_0(\vect{k})$, and the two-particle interactions $\mathcal{V}_0$ are given in \Cref{a:s:mean_field}.
        The off-diagonal bilinear part $\Gamma_0$ is zero after the self-consistent solution of the mean-field decoupling.

        After the mean-field decoupling and the CST flow the effective Hamiltonian is
        \begin{align}
            \label{eq:effHamil}
            \mathcal{H}_{\mathrm{eff}}= E + \sum_{\vect{k}} \disp(\vect{k}) 
	    			\big(\norord{ \alp*{\vect{k}}\alp{\vect{k}}} + 
	    			\norord{\bet*{\vect{k}}\bet{\vect{k}}}\big) + \mathcal{V}.
        \end{align}
        with the dispersion $\disp(\vect{k})$ of the quasi-particles in the new basis and the quasiparticle conserving interaction term $\mathcal{V}$.

        \subsection{Inverse Participation Ratio} \label{ss:ipr}
        In \Cref{ss:boundstates} we determine for each of the four bound states at which value of the anisotropy $\lambda$ it enters the two-magnon continuum, i.e., where it ceases to exist.
        This merging into a continuum generically occurs slowly in the sense that the slope of the lower band edge of the continuum and of the energy of the bound state are the same \cite{Uhrig_1996,Uhrig_1996_err,Zhitomirsky_2013}. 
        Hence, the parameters where	the bound state ceases to exist are extremely difficult to determine numerically. 
        Therefore, we use another criterion, namely the locality of the bound state	in its relative coordinate.
        A true bound state is local in the sense that it has a finite extension.
        If the extension is given by the system size no binding occurs anymore.
        A good measure of the locality of a state is the inverse participation ratio (IPR) \cite{Kramer_1993,Malki_2019}
        \begin{equation}
	    	\label{eq:ipr}
            I = \sum_{\vect{r}}  | \Psi(r) |^4 = \sum_{\vect{r}}  | \rho(r) |^2 .
        \end{equation}
        with density $\rho(r)$ at site $r$ of the normalized wave function $\Psi(r)$.
        For an extended state the IPR $I$ scales proportional to $1/L^2=1/N$ if $L$ is the number of sites in one direction in two dimensions,
        This is easily seen in the limiting case of $\rho(r)=1/N$ where $I=1/N$ holds.
        The other limiting case is a completely local state with $\rho(0)=1$ and zero elsewhere so that $I=1$ holds.
        Hence one has to determine whether the IPR vanishes $\propto 1/N$ for $N\to\infty$ or stays finite in this limit.
    
        \subsection{Flow equation and residual off-diagonality}
	    \label{ss:flow-rod}
    
        The flow equations are a set of ordinary differential equations (ODE) for the coefficients of the contributing monomials of operators which depend on wave vectors. 
        We solve these ODEs numerically for a finite lattice with a standard Dormand-Prince 5 method as implemented by the \textit{odeint} project \cite{Ahnert_2011} up to linear lattice size $L=22$.
        In order to avoid the accumulation of numerical errors and to enhance the performance we use the point group symmetries in the MBZ as depicted in \Cref{fig:mbz_mesh}.
        In addition, we use an unambiguous symmetrized representation of the coefficients.
        We stop the flow when the residual off-diagonality (ROD), i.e., the square root of the sum over all squared entries of the generator $\eta$ \cite{Fischer_2010}, has dropped to values below $10^{-6}J$.  

        \subsection{Discretization of the magnetic Brillouin zone} \label{ss:discrete}
        One has to pay special attention to the wave vector $\vect{k} = (0,0)$ at $\lambda \lesssim 1$.
        The gapless mean-field solution \eqref{eq:startHamil} at the isotropic point displays diverging coefficients whenever one of the wave vectors in the quadrilinear monomials vanishes, see for instance \refcite{Uhrig_2013}.
        Hence, this solution cannot be used for a discretized MBZ. Note that these divergences are integrable in the thermodynamic limit where the wave vector dependence is continuous.
        So they do not constitute a physical, but rather a numerical issue. 
        For $\lambda \to 1$ and especially for small $L$ we encountered signs of problematic divergences in form of a non-converging flow.
        One way to circumvent this problem consists in using a reciprocal lattice where $\vect{k} = (0,0)$ does not occur, for instance by using odd values of $L$ with antiperiodic boundary conditions as illustrated in right panel of \Cref{fig:mbz_mesh}.
        In real space, the antiperiodic boundary conditions are achieved by giving spin flip terms across the boundary of the system's cluster a minus sign.
        In this case, the cluster is a rhombus with side length of $\sqrt{2}L$. 
        In the following, we denote lattice discretizations sampling $\vect{k} = (0,0)$ by $\mbzO$ and discretizations not sampling $\vect{k} = (0,0)$ by $\mbz$.
        We also implemented the approach used by Powalski et al.~\cite{Powalski_2015, Powalski_2018} where coefficients are set to zero, for which at least one wave vector vanishes, i.e., they are neglected.
        The results extrapolated to $L \rightarrow \infty$ agree well with the results presented below, but they show a significantly slower convergence in $1/L$.
        For this reason we do not consider them any further. 

        Since the approach $\mbz$ excludes the wave vector at which the spin gap $\Delta = \disp(\vect{k}=(0,0))$ can be read off by construction, the gap value must be retrieved otherwise.  
        To this end, we adapt the fit
        \begin{align}
            \label{eq:dispfit}
            \disp(k)^2 = \sum_{m=0}^M A_m \cos(m k), 
        \end{align}
        used in Ref. \cite{Okunishi_2001} for a one-dimensional model to our two-dimensional case.
        We fit the discrete dispersion values obtained by CST along the line from $(0,0)$ to $(\pi,0)$ in order to determine the coefficients $A_m$.
        A very fast convergence of these coefficients is found for increasing $M$ so that we can interpolate the dispersion very reliably in the whole \mbz.

        \subsection{Determination of correlation lengths}\label{ss:det-correlationlength}
        As mentioned in Sec.~\ref{ss:mf}, we perform the mean-field decoupling to determine the initial conditions in the thermodynamic limit, while we solve the flow equations for a discretized lattice as discussed in the previous section.
        Hence, our results combine elements of calculations in the thermodynamic limit with calculations on a finite lattice.
        In order to judge if the discretization in the MBZ induces finite-size effects, we calculate the ground-state correlation length $\xi$.
        Whenever it is significantly larger than the linear length scale of the system, finite-size effects are likely to occur.
        We expect the correlation length $\xi$ to be the same in all directions for $\xi\gg a$ where $a$ is the lattice constant.
        For $\lambda \rightarrow 1$, $\xi$ diverges. 
        Thus, it is sufficient to perform the same one-dimensional fit for a single direction as in \Cref{ss:discrete}, \Cref{eq:dispfit} and calculate $\xi$ from the one-magnon dispersion $\disp$ via
        \begin{align}
            \omega(\im \kappa) = 0\quad \text{with} \quad \mathrm{Re}=1/\xi
        \end{align}
        adapted from \refcite{Okunishi_2001}.


            


        \subsection{Extrapolation in the linear system size $L$}\label{ss:extrapol-linear}
\begin{figure}[!ht]
            \centering
            \includegraphics[width=\columnwidth]{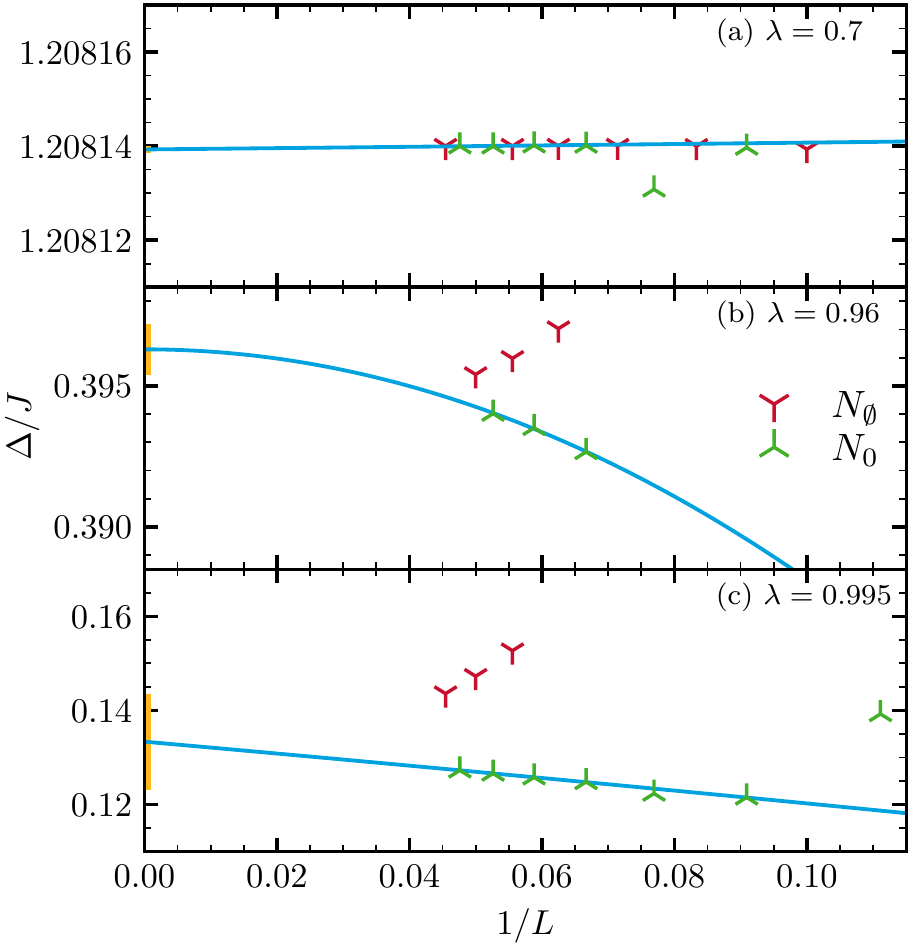}

            \caption{Generic extrapolations in $1/L$ for $\lambda=0.7$, $\lambda=0.96$, and $\lambda=0.995$. For small $\lambda$, calculations on the meshes \mbz\, and \mbzO converge, see panel (a).
            At higher values of $\lambda$, see panel (b) and (c), values from $\mbz$ are monotonically increasing while values from \mbzO{} are monotonically decreasing. 
            For the whole range, a monotonic quadratic fit for the  $\mbz$ data is used to determine the values for $L \rightarrow \infty$. 
            The error estimate depicted in yellow results from the difference between the values from $\mbz$ and $\mbzO$ for the highest reached value of $L$. }  
            \label{fig:convergence}
        \end{figure}

        For all results shown an extrapolation in $1/L\to 0$ is performed.
        For small values of $\lambda$ up to $\lambda\approx 0.8$, the computed values do not show any relevant finite-size effects.
	    There is no discernible difference between the two kinds of mesh $\mbz$ and $\mbzO$.
        Hence, all extrapolating schemes yield essentially the same result, see panel (a) in \Cref{fig:convergence}.
        For $\lambda > 0.8$, we observe two trends for increasing $L$, see panels (b) and (c) in \Cref{fig:convergence}.
        First, the values for $\mbzO$ decrease monotonically decreasing while those for $\mbz$ increase monotonically.   
        Second, the values from the mesh $\mbz$ show an almost linear relation as function of $1/L$ for $\lambda \lesssim 1$.
        The values we provide as the estimate of the extrapolation are the results of monotonic quadratic fits 
        \begin{align}
                x(1/L) =  a + b( 1/L+c )^2
        \end{align}
        for the mesh $\mbz$ with $c\ge0$, evaluated at $1/L = 0$.
        The error estimate results from the difference between the values from the meshes $\mbz$ and $\mbzO$, see \Cref{fig:convergence}. For the correlation length $\xi$ a linear extrapolation in $1/L$ was used; the error estimate remains the same.

        The locality of bound states is measured by the IPR given in \Cref{eq:ipr}. 
        This quantity is extrapolated linearly in $1/N$ which works remarkably well yielding positive values for bound states and values in the close vicinity of zero otherwise.
        
    \section{Results}
        \label{s:results} 
    \subsection{Convergence}  \label{ss:convergence}
        Generally, flow-equation approaches based on the quasi-particle conserving generator transform the original Hamiltonian to a block-diagonal form, where each block describes the action of the effective Hamiltonian on the subspace of a given, constant number of quasi-particles, here magnons, determined by $Q$.
        If, however, this quasi-particle picture breaks down, for example by a second-order phase transition or by level crossings of modes with a different number of quasi-particles, this separation into blocks is not possible anymore. 
        As a consequence, the flow does not converge and no description in terms of the chosen quasi-particles is obtained. 
        This feature makes sense because it reflects that the physics can no longer be captured in terms of the particular choice of quasi-particles.
        Conversely, if the flow converges this indicates that the underlying quasi-particles picture does not break down and that there are no inter-block level crossings in the chosen truncation scheme. 
        Hence, the description in terms of the chosen quasi-particles reflects the underlying physics correctly.

        The flow of the CST calculations with the truncation based on scaling dimension converges for both types of meshes $\mbz$ and $\mbzO$ for all $\lambda$.
        Only for $\lambda \ge 0.99925$ and the mesh $\mbzO$ including $\vec{k} = (0,0)$ we find a divergent flow, that can be traced back to numerical artifacts, see Sect.\ \ref{ss:discrete}.
        In general, an analysis of the ground-state energy, the magnon dispersion, and bound states in the two-magnon block is feasible and well-founded. 

    \subsection{Ground-state energy}
    \label{ss:ground_state}
\begin{figure}[htb]
            \includegraphics[width=\columnwidth]{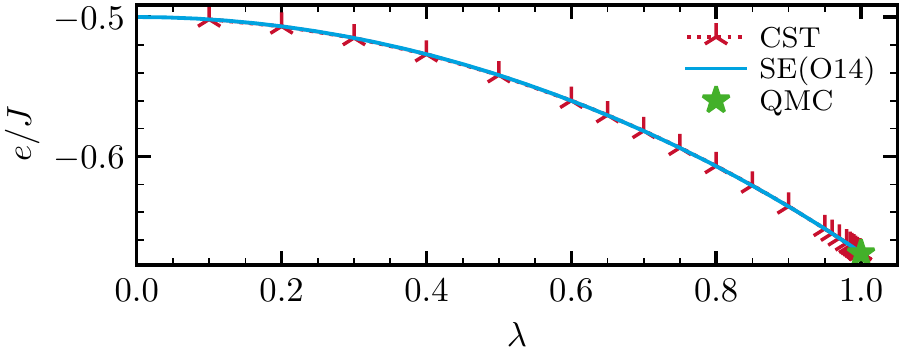}

            \caption{Ground-state energy per site $\gsPerSite$ as a function of $\lambda$. Red crosses are CST data which are compared to results of a series expansion(SE) \cite{Zheng_1991} drawn as blue line; the green star marks the QMC results for the isotropic Heisenberg model \cite{Sandvik_1997}. As expected from second-order perturbation theory the ground-state energy displays a monotonically decreasing behavior with negative curvature.}
            \label{fig:gs:whole_range}
        \end{figure}

        The first quantity of the XXZ-model we inspect is the ground-state energy per site $\gsPerSite (\lambda) = \gs(\lambda) / ( 2L^2 )$. It is shown in \Cref{fig:gs:whole_range} as computed by CST, series expansion up to order 14 in $\lambda$ \cite{Zheng_1991}, and QMC calculations \cite{Sandvik_1997}.
        Starting from the classical N\'eel state in the Ising limit,  increasing $\lambda$ induces more and more quantum fluctuations resulting in a ground-state energy which continuously decreases. 
        The results of the CST are identical to the series expansion results within line width for the whole $\lambda$ range.
        For $\lambda = 1$, the CST result ($\gsPerSite =\num{-0.669 38}\,J$) agrees well with QMC calculations ($\gsPerSite = \num{-0.669 436(7)}\,J$), and the extrapolated series expansion results ($\gsPerSite = \num{0.6693(1)}\,J$).
        We recall, however, that the self-consistent spin-wave theory at $\lambda =1$ already yields $\gsPerSite =\num{-0.670 421}\,J$, fairly close to the above values in spite of its mean-field character. 
        This shows that the ground-state energy is a very robust quantity, accessible by many approaches.
        This is different for the dispersion which we consider next.

    \subsection{Dispersion}
    \label{ss:dispersion}
\begin{figure}[htb]
            \includegraphics[width=\columnwidth]{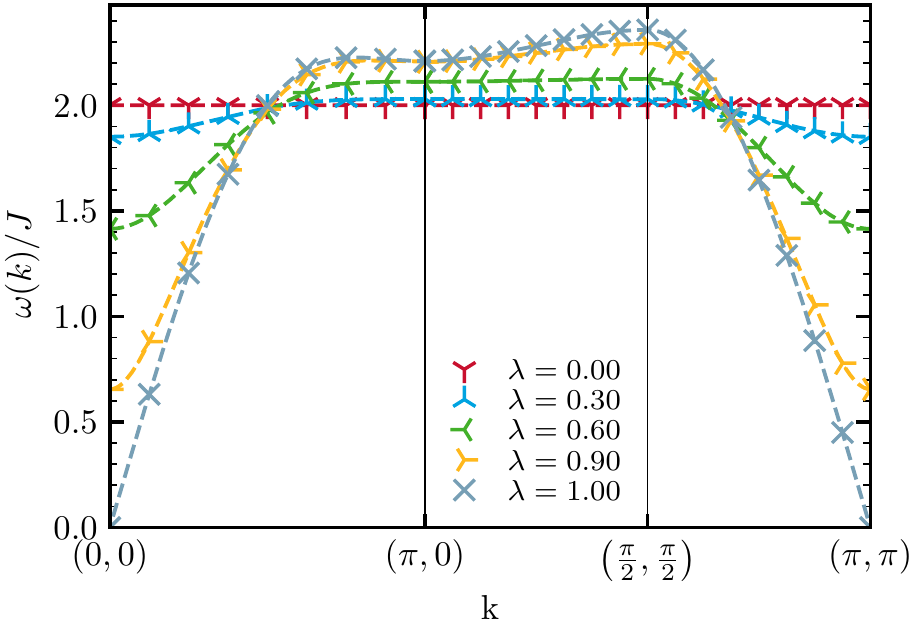}
            \caption{CST result of the dispersion $\disp(\vect{k})$ for $\mbzO$ and $L=16$ for various $\lambda$. For $\lambda=0$, the XXZ model is the Ising model with a flat dispersion while it is the antiferromagnetic Heisenberg model for $\lambda=1$ with a gapless spectrum and a distinct roton minimum at $\vect{k}=(\pi,0)$ \cite{Powalski_2015,Verresen_2018}. For \mbox{$0 < \lambda < 1$} the CST results interpolate smoothly between these two limits.}
            \label{fig:disp:full}
        \end{figure}

        The results of the CST for the dispersion along a high-symmetry path through the MBZ for various $\lambda$ and $L=16$ are shown in \Cref{fig:disp:full}.
        At $\lambda = 0$, the XXZ model is reduced to the antiferromagnetic Ising model with a completely flat dispersion $\disp(\vect{k})=2J$.
        For $\lambda = 1$, the XXZ model coincides with the antiferromagnetic Heisenberg model which was already studied using the CST method in Refs.\ \cite{Powalski_2015,Powalski_2018}.
        At this isotropic point, the dispersion features gapless Goldstone bosons and a distinct minimum at $\vect{k} = (\pi,0)$, called the roton minimum. 
        For $0 < \lambda < 1$, a continuous evolution between these two limits takes place. Spin gap and roton minimum are discussed below.

        \subsubsection{Spin gap}
        \label{sss:gap}
\begin{figure}[htb]
                \centering
                \includegraphics[width=\columnwidth]{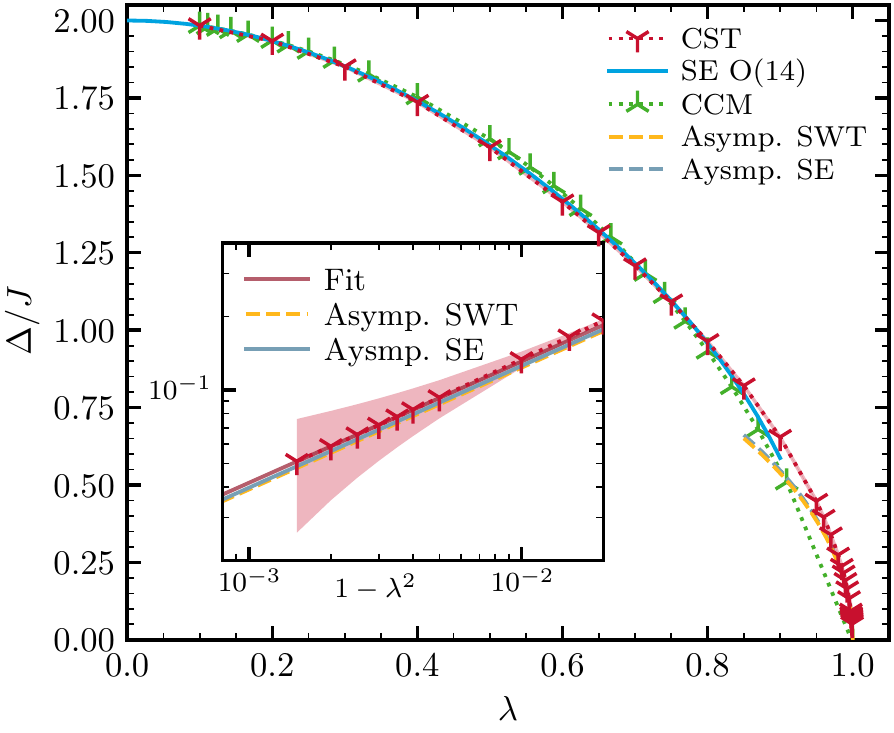}
                \caption{CST results for the one-magnon gap $\gap = \disp(\vect{k} = 0)$ of the XXZ model for $0 \le \lambda \le 1$ compared to third order spin wave theory (SWT) \cite{Hamer_1992, Syromyatnikov_2010}, data from Coupled Cluster Method (CCM) \cite{Bishop_2017}, results from series expansion (SE) about the Ising limit \cite{Zheng_2005, Dusuel_2010}, and the critical power law extracted from SE \cite{Zheng_2005}. A quantitative comparison of the critical behavior is shown in the inset underlining very good agreement.}
                \label{fig:disp:gapfull}
            \end{figure}

            \Cref{fig:disp:gapfull} shows the extrapolated CST results for the one-magnon gap $\gap$ compared with previous results.
            For small $\lambda$, we compare to high-order series expansion calculation \cite{Zheng_2005, Dusuel_2010} and to data obtained by the Coupled Cluster Method (CCM) \cite{Bishop_2017}.
            For large $\lambda \lesssim 1$, the asymptotic power law in the critical region derived from series expansion results \cite{Zheng_2005} and spin-wave theory \cite{Hamer_1992} are presented.
            The CST results interpolate smoothly between these two limits, capturing both limits quantitatively.
            The inset in \Cref{fig:disp:gapfull} shows a double logarithmic plot of the gap as a function of $1-\lambda^2$ in the critical region, i.e., for $\lambda\rightarrow1$.
            The last data point for $\mbzO$ and $L=22$ is $\lambda = 0.99925$; for higher values no reliable convergence was achieved. 
            We attribute this to the fact that for values of $\lambda$ even closer to the isotropic point the cluster size $L$ is not large enough to capture the relevant physics, see also \Cref{fig:disp:corrlength} in \Cref{sss:corrlength}.
            Our error estimate (see also inset of \Cref{fig:disp:gapfull}) grows for larger values of $\lambda$ implying larger correlation lengths in the vicinity of the isotropic point.
            For a quantitative analysis, we fit 
            \begin{eqnarray}
                \Delta(\lambda \approx 1) = c \left(1 - \lambda^2 \right)^{\nu z},
            \end{eqnarray}
            which is shown in the inset.
            We obtain $c = \num{1.30(2)} $ and $\nu z=\num{0.498(2)}$ which agrees within 5\% with the value $c=\num{1.265(2)}$ obtained from series expansion \cite{Zheng_2005} if the critical exponent $\nu z=1/2$ was fixed.
            For fit with fixed exponent of $\nu z=1/2$ in our results we see a slight increase to $c=\num{1.313(1)}$ which still agrees within 5\% to the literature.
            The agreement between all approaches is very good for almost all values of $\lambda$.
            For smaller values of $\lambda$ the various approaches agree excellently.
            Only close to the vanishing of the gap, the CCM data deviates from the other results which is likely to be linked to the attainable cluster sizes.
            The agreement between the extrapolated series expansion and the CST results is remarkable since the latter are obtained on a finite lattice of moderate size $L\approx 20$ and extrapolated to the thermodynamic limit without imposing a certain power law. 
            This finding corroborates the applicability of CST in momentum space to describe phase transitions with closing gaps.
		 
        \subsubsection{Correlation length $\xi$ }
        \label{sss:corrlength}
\begin{figure}[htb]
                \centering
                \includegraphics[width=\columnwidth]{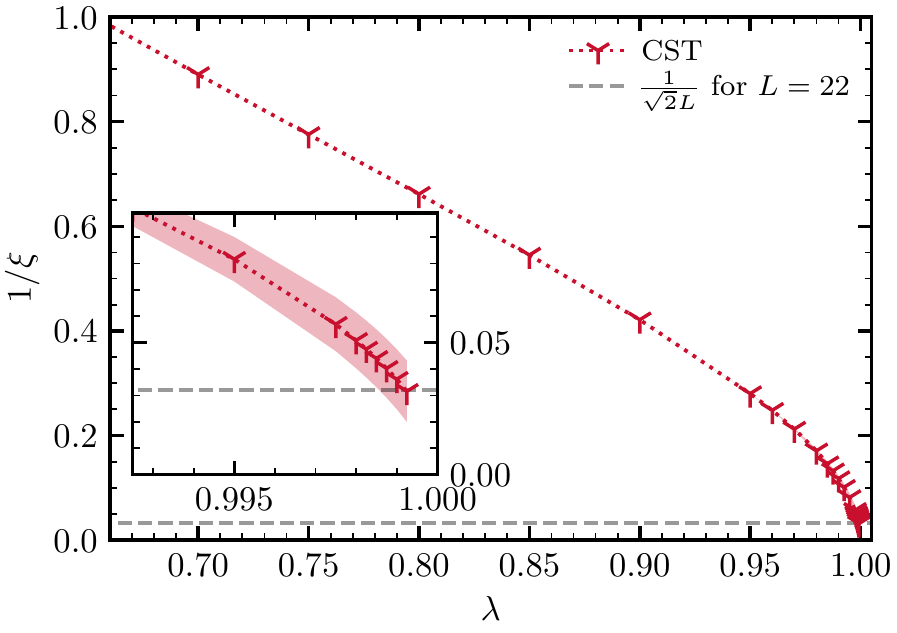}
                \caption{
                    CST results for the ground-state correlation length $\xi$ for $0.7 \le \lambda \le 1$. 
                    A comparison of the correlation length with the shortest wrap around $\sqrt{2}L$ in the largest cluster with $L=22$ (dashed line) is shown.
                   }
                \label{fig:disp:corrlength}
            \end{figure}

            \Cref{fig:disp:corrlength} shows inverse correlation lengths $1/\xi$ along the used fit-axes from $(0,0)$ to $(\pi,0)$ for different values $\lambda$.
            Note that we use units $1/a = 1$, since the lattice constant $a$ is set to $1$.
            Additionally, we show a horizontal dashed line for the relevant length scale of our largest system $L=22$, namely the shortest occurring wrap-around $\sqrt{2}L$, in this case, in diagonal direction.
            In the inset of \Cref{fig:disp:corrlength} a zoom to the domain $\lambda > \num{0.993}$ is shown.

            For values of $\lambda$ up to $\num{0.995}$ we find finite correlation lengths, well below $\sqrt{2}L$.
            For $\lambda > \num{0.995}$, $1/\xi$ starts to vanish, which is expected since the system at $\lambda = 1$ is gapless with infinite correlation length.
            The error bar for values of $\lambda > 0.9985$ crosses the threshold $1/\sqrt{2}L$ arising from the finite system size.
            Additionally, for the largest achieved value of $\lambda = 0.99925$, $1/\xi$ itself approaches $1/\sqrt{2}L$.

            We conclude that for $\lambda \le \num{0.995}$, finite-size effects are not expected. 
            For $\lambda \ge 0.9985$ the finite size of the system might play a noticeable role.
            However, $\xi$ never exceeds these relevant length scales dramatically.
            Additionally, the extrapolated results for closing of the gap are in agreement with the literature and the CST results for the Heisenberg model in Ref. \cite{Powalski_2015,Powalski_2018} performed for $L=16$ are consistent with results in the literature.
            Hence, it appears that the influence of the finite size is not severe in the CST approach.
            
            The error estimate from $\lambda = 0$ to $\lambda = 0.95$ lies below $\SI{0.1}{\%}$, below $\SI{1}{\%}$ for $\lambda \le 0.97$, below $\SI{5}{\%}$ for $\lambda \le 0.99$ and it is larger than $\SI{10}{\%}$ for $\lambda \ge 0.995$.

        \subsubsection{Roton mode}
			\label{sss:roton}
				
\begin{figure}[htb]
                \centering
                \includegraphics[width=\columnwidth]{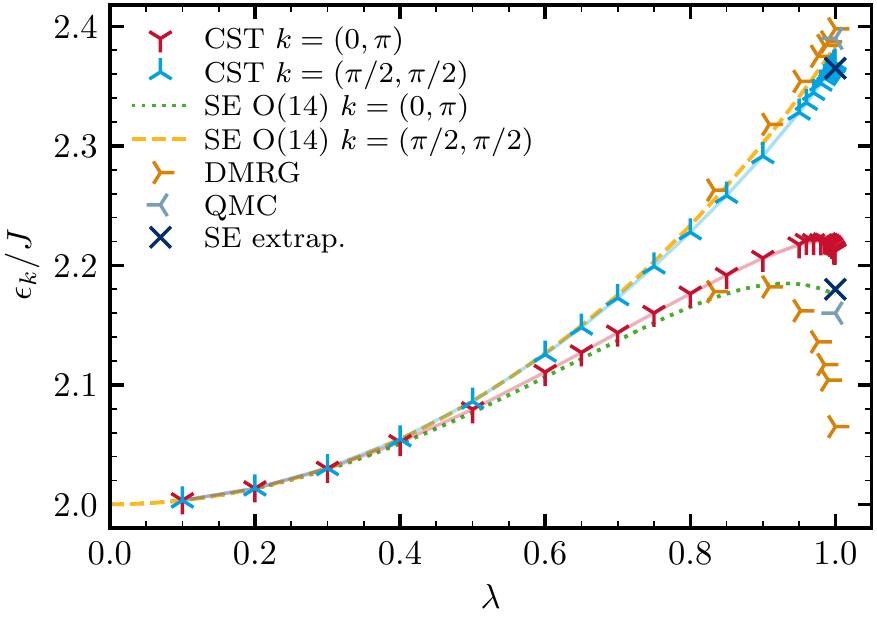}
                \caption{CST results for the roton mininum $\vect{k} = (\pi,0)$ and the dispersion maximum $\vect{k} = (\pi/2,\pi/2)$ for $0 \le \lambda \le 1$ compared to data from series expansion (SE) \cite{Zheng_2005}, DMRG \cite{Verresen_2018}, and QMC \cite{Sandvik_2001}. All methods agree well for dispersion maximum. 
                For the roton mininum, all methods predict an inflection point for $\lambda \gtrsim 0.8$. The values for the depth of the roton mode differ slightly.}
                \label{fig:disp:RotonMinMax}
            \end{figure}

            \Cref{fig:disp:RotonMinMax} depicts the CST results for both the roton minimum and the dispersion maximum as a function of $\lambda$. 
            Additionally, results from series expansion \cite{Zheng_2005} and DMRG \cite{Verresen_2018} are shown. 
            Also, QMC data \cite{Sandvik_2001} are available for the isotropic point.
            First, we focus on the dispersion maximum at $\vect{k} = (\pi/2,\pi/2)$ which displays a monotonic increase in $\lambda$ for CST, the series expansion and the DMRG data. All data agrees very well.
            At the isotropic point, similar values are obtained: CST (\num{2.36972(5)}\,$J$), series expansion (\num{2.385(1)}\,$J$), QMC(\num{2.39}\,$J$), and DMRG(\num{2.40}\,$J$).            
        
            For the roton minimum, CST, series expansion, and DMRG show qualitatively similar results: first the dispersion increases and for larger $\lambda \gtrsim 0.8$ an inflection point occurs. This inflection point is consistent with the observation in Ref. \cite{Powalski_2015,Powalski_2018} that the hybridization between the one-magnon state and the three-magnon scattering states causes the dip due to level repulsion.
            This mechanism matters more and more for higher $\lambda$ when the dispersion and the continuum approach each other.
            The numerical values for the roton dip for $\lambda=1$ are \mbox{\num{2.211(3)}\,$J$} for the CST, \num{ 2.18(1)}\,$J$ for series expansion, \num{2.16}\,$J$ for QMC, and \num{2.06(1)}\,$J$ for DMRG.
            Hence, these values are reasonably close, but further spread than the maximum values.
            We assume that the SE and QMC results reflect the true value best while CST is still a bit too high because of truncation beyond the quartic monomials. 
            The DMRG data results from a cylindrical geometry with fixed linear extension $L=10$.
            A comprehensive finite-size scaling has therefore could not been performed.
            All in all, we conclude that a consistent description for the high-energy properties of the dispersion of the XXZ model is reached by CST except for a slight deviation of about $\SI{2}{\percent}$ at the roton minimum.
        \subsection{Bound states and two-magnon continuum}
        \label{ss:boundstates}
        A particular asset of the systematic basis changes by means of CST is that bound states can be directly addressed and computed.
        The conservation of the particle number, here the magnon number, is achieved by the basis change. 
        It allows one to compute the bound states in the two-particle sector. 
        Recently, even three-particle bound states could be addressed in antiferromagnetic spin ladders which are induced by three-particle irreducible interactions, i.e., by hexatic terms in second quantization \cite{Schmiedinghoff_2022}.
        Here, we focus on two-magnon bound states as established perturbatively for not too large values of $\lambda$ \cite{Hamer_2009,Dusuel_2010}.

        For bound states to be infinitely long-lived they may not decay into scattering states. 
        Hence, their energy may not overlap with the corresponding continuum; it should stay below the lower boundary of the continuum. 
        Generically, the energetically lowest lying bound states occur at total momentum zero $\vec{K}=(0,0)$ corresponding to the bound state at rest.
        This is confirmed by series expansions \cite{Hamer_2009,Dusuel_2010}.
        Hence, we compute four bound states for zero center-of-mass momentum $\vect{K}=\vect{k}_1+\vect{k}_2=(0,0)$.
        In addition, the strongest binding occurs for zero total $S^z$, i.e., between an $\alpha$- and a $\beta$-magnon living on the A- and B-sublattice, respectively. 
        Again, this is confirmed perturbatively.

\begin{figure}[htb]
                \centering
                \includegraphics[width=\columnwidth]{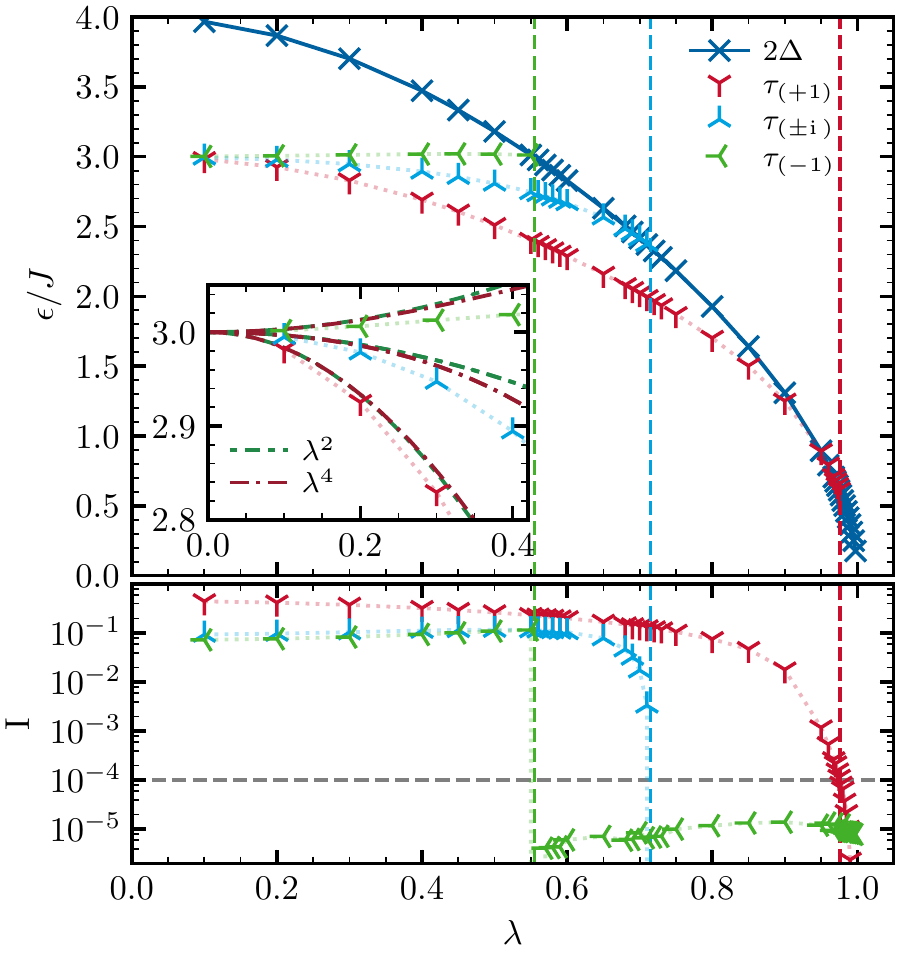}
                \caption{The upper panel shows the energies of the four magnon-magnon bound states $\boundstates$ calculated by CST as a function of the anisotropy $\lambda$ and the corresponding lower edge $2\Delta$ of the two-magnon continuum.
                The inset compares data from truncated CST to data from perturbative CUT up to terms $\propto\lambda^4$ \cite{Dusuel_2010}.
                One discerns a certain small deviation in quadratic order.
                In the lower panel, the inverse participation ratio IPR of the four bound states is plotted; only three curves are shown because the eigen wave functions of the two degenerate bound states are complex conjugates with the same IPR $I$.
                The vertical dashed lines mark the decay points for the corresponding bound state determined by a crossing of $I$ with the threshold $I=10^{-4}$ (horizontal dashed line).}
                \label{fig:2magnon:Full}
        \end{figure}

        In order to distinguish the four states and to avoid numerical inaccuracies we exploit the point group symmetry of rotations $R$ by $90^\circ$ about any site of the lattice; the effective Hamiltonian is block-diagonal within each eigen subspace of $R$.
        Four eigen values of $R$ are possible and do occur: $\pm 1$ and $\pm i$. Since the XXZ model is time-reversal invariant the imaginary eigen values are degenerate and their complex eigen vectors are complex conjugates.
		The numerical diagonalization within each eigen subspace of $R$ yields the energies of the bound states as depicted in \Cref{fig:2magnon:Full}.  

        Starting in the Ising limit, we find four bound states $\boundstates$ below the lower edge of the two-magnon continuum. 
        For increasing $\lambda$ they are absorbed in the continuum one after the other before the isotropic Heisenberg model is reached.
        We determine the decay point where each bound state merges with the continuum by computing and extrapolating its IPR as described in \Cref{ss:ipr} and \Cref{ss:extrapol-linear}.
        We recall that a finite IPR indicates binding while its vanishing indicates delocalization, i.e., the bound states ceases to exist. 
        The results are displayed in the lower panel of \Cref{fig:2magnon:Full}.
        It is advantageous to determine the merging into the continuum from the vanishing of the IPR because this vanishing occurs as very rapid drop by many orders of magnitude while the crossing of the binding energies with the lower continuum edge is an intersection with small, rigorously even vanishing, angle.
        We choose a numerical threshold for the IPR to distinguish between a bound state and a delocalized state in the continuum and set it to $\num{1e-4}$, i.e., only for \mbox{IPR $>\num{1e-4}$} we deduce binding. 
        This value is the typical value we found for the IPR of generic states well within the continuum, extrapolated to the infinite thermodynamic limit $N\to\infty$.
        The absorption points are marked as dashed vertical lines.

        We find that $\boundstate{-1}$ disappears for $\lambda\approx \num{0.565(5)}$ which is a deviation of $\approx\SI{5}{\percent}$ to $\lambda\approx \num{0.5401(1)}$ from the pCUT \cite{Dusuel_2010}. 
        The merging of  $\boundstate{\pm i}$ occurs at $\lambda\approx \num{0.72(1)}$ in CST; no results were given for these bound states in Ref.\ \cite{Dusuel_2010}. 
        For the curves for $\boundstate{\pm i}$ in the inset we analyzed the hopping amplitudes given in \refcite{Dusuel_2010} to calculate the series up to order $\lambda^4$.
        For $\boundstate{+1}$ we find $\lambda\approx \num{0.975(5)}$ with a deviation of only $\approx\SI{0.6}{\percent}$ to the extrapolated pCUT value of $\lambda\approx \num{0.97}$. 
        We conclude that for almost all values $\lambda\in[0,1]$ at least one bound state exists.
        But even the lowest bound state $\boundstate{+1}$ dives into the continuum very close to the isotropic Heisenberg model.
        We note that this finding is very similar to the decay of bound states in the XXZ model on coupled two-leg ladders. 
        Here QMC simulations reveal the decay of the lowest bound state at $\lambda\approx \num{0.96}$ for different inter-ladder couplings \cite{Ying_2019}.
        
        Comparing the CST result with the results from the perturbative series near the Ising limit, see inset of \Cref{fig:2magnon:Full}, we discern a sizable deviation in order of $\lambda^2$ for all bound states, in contrast to what we found for the ground-state energy and the dispersion.
        These deviations are a consequence of the truncation scheme neglecting hexatic operators.
        Indeed, one finds that for a perturbative CST in second order of $\lambda$ in position space, coefficients for quartic interactions of the type $\abos*{i}\abos{i}\bbos*{j}\bbos{j}$ are coupled to hexatic terms during the flow.
        Hence, these contributions are not considered rigorously in the CST as implemented in this paper.
        Therefore, the quantitative analysis for properties at intermediate values of the anisotropy has to be taken with a grain of salt.
        Nevertheless, we observe that the series expansion and the CST results show very similar behavior.
        In particular, the results close to the Heisenberg point agree very well.
        Obviously, the benefits of the truncation based on the scaling dimension outweighs the caveat of not capturing all perturbative terms exactly in the Ising limit.
        This observation holds an attractive promise for future application of the CST combined with a truncation based on scaling.

		
    \section{Summary and Outlook}
    \label{s:summary}
		
        We extended the systematic basis change by a continuous similarity transformation as applied to the spin isotropic Heisenberg model in two dimensions in Refs.\ \cite{Powalski_2015,Powalski_2018} to the two-dimensional XXZ model.
        This model displays the transition from a spontaneously broken discrete symmetry with finite energy gap to a spontaneously broken continuous symmetry with vanishing energy gap. Starting point of our calculation was the Hamiltonian in self-consistent spin wave theory based on the Dyson-Maleev representation of the spin operators. The calculation is performed by numerically integrating the flow equations in the prefactors of terms in second quantization. To this end, we discretized the magnetic Brillouin zone in various ways and checked that the results converged within small error bars. 
        In addition, the proliferation of terms was limited by a truncation based on the scaling dimension. 
        We kept all terms of quadratic and quartic form, but neglected hexatic and higher terms after normal-ordering them with respect to the magnon vacuum.

        We verified that this approach can be applied to the paradigmatic transition of the two-dimensional XXZ model including the vanishing of the spin gap. 
        The ground-state energy, the one-magnon dispersion as well as the two-magnon interaction are derived and computed. The dispersion displays the expected appearance of the roton minimum and agrees well with all available previous results. 
        We successfully computed the two-magnon bound states at zero momentum of the center-of-mass in the longitudinal channel, i.e., for vanishing total $S^z$ component. 
        Four bound states are identified of which two with complex eigen wave functions are degenerate. The latter states have not been addressed so far. 
        Upon approaching the isotropic Heisenberg model they merge successively with the two-magnon continuum. 
        The lowest bound state is rotationally invariant in space, i.e., with respect to rotations by $90^\circ$, and vanishes only very close to the isotropic point in nice quantitative agreement with previous perturbative results \cite{Dusuel_2010} within less than one percent. 
        This agreement is even more remarkable because the energies of the bound states show a larger deviation of the order of five percent to the perturbative results for intermediate values of $\lambda$.
        We trace this effect back to the truncations at beyond quartic level. 
        The degree of binding and the delocalization upon diving into the continuum are determined reliably by the inverse participation ratio.
 
        The above sketched achievements hold the promise that the systematic CST with truncation based on the scaling argument can be applied to many more models which up to date have not been studied in great detail. 
        We emphasize that the closing and opening of gaps can be captured as well as the occurrence of binding phenomena.
        Models suggesting themselves comprise the bilayer model of two stacked square lattices which displays a transition to from the long-range ordered phase with magnons to triplons on the interlayer bonds upon increasing interlayer coupling \cite{wang06,lohoefer15}.
        The frustrated $J_1$-$J_2$ model on the square lattice is another intriguing candidate to be investigated its two limits displaying long-range order \cite{uhrig09a,jiang12b,gong14,richt15,morit15,ferra18}. 
        Antiferromagnets displaying non-collinear long-range order represent another wide field of application. 	

    \begin{acknowledgments} 
        We thank Frank Pollmann and Ruben Verresen for providing the DMRG results for the roton modes and Stefan Wessel and Andres Sandvik for useful discussions. 
        We gratefully acknowledge financial support by the German Research Foundation (DFG) in projects UH 90-13/1 (GSU), UH 90-14/1 (DBH) and SCHM 2511/13-1 (MRW/KPS).
    \end{acknowledgments}

    \bibliography{bibliography.bib}
		
    \begin{appendix} 
		
      \section{Mean-field decoupling in real space}
      \label{a:s:mean_field} 
      Here we provide the explicit expressions for each step of the derivation of the starting point of the CST, i.e., for determining the self-consistent mean-field solution. Apart from the spin anisotropy $\lambda$, the calculations are analogous to the ones in Refs.\ \cite{Powalski_2015, Powalski_2018, Uhrig_2013}.
      The Hamiltonian in  Dyson-Maleev representation reads
      \begin{align}
            \mathcal{H} &= J \sum_{i \in \Gamma_A, \delta}  \left[ - S^2 \nonumber \right.\\
                        &+ S \left( \abos*{i}\abos{i} +\bbos*{i+\delta}\bbos{i+\delta} + \lambda \abos{i} \bbos{i + \delta} +  \lambda \abos*{i} \bbos*{i + \delta} \right)\\
                        \nonumber &- \left. { \abos*{i} \abos{i} \bbos*{i+\delta}  \bbos{i+\delta} }
                            -\frac{\lambda}{2} { \abos*{i} \abos{i}  \abos{i} \bbos{i+\delta} }
                         -\frac{\lambda}{2} { \abos*{i} \bbos*{i+\delta}  \bbos*{i+\delta} 
														 \bbos{i+\delta} } \right],
      \end{align}
      where we switched the sum over all adjacent pairs$\langle i,j\rangle$ in Eq.\ \eqref{eq:hamiltonian} to a sum over all sites in sublattice A, i.e.,  $\Gamma_A$, and their nearest neighbors on the square lattice reached via the distances $\delta$. 
      The first line represents the classical energy, the second the bilinear one-magnon terms and the last line the quadrilinear two-magnon terms.			
      Next, we perform the standard mean-field decoupling \cite{Takahashi_1989} in real space 
      \begin{align}
                b^{(\dagger)}_i \tilde{b}^{(\dagger)}_j =  
								\norord{ b^{(\dagger)}_i \tilde{b}^{(\dagger)}_j} +  
								\cexpval{ b^{(\dagger)}_i \tilde{b}^{(\dagger)}_j },
      \end{align}
      where $\norord{...}$ indicates normal-ordered operators and $\cexpval{ ... }$ are the expectation values in the mean-field ground-state.
      For quadrilinear terms we apply the corresponding relations based on Wick's theorem.

      To make the resulting expression more concise, we use exact relations and conservation laws. 
      First, the Hamiltonain is invariant under the exchange of the A and B sublattice. For the Dyson-Maleev bosons this translates into a complex conjugation of the prefactors and swapping $a_{i}^\dag\leftrightarrow b_{i+\delta}$ and $a_{i}\leftrightarrow b_{i+\delta}^\dag$.
      We define the expectation values
      \bes
      \label{eq:n_Delta}
      \begin{align}
                n & \coloneqq \cexpval{ \abos*{i} \abos{i}} 
								= \cexpval{ \bbos*{i+\delta} \bbos{i+\delta}}\\
                \Delta & \coloneqq \cexpval{ \abos{i} \bbos{i+\delta} } 
								= \cexpval{ \abos*{i} \bbos*{i+\delta}}, 
      \end{align}
      \ees
			which are determined self-consistently below.
      
      Second, the Hamiltonian \eqref{eq:hamiltonian} conserves the total spin component
			\be
                S^z_{\rm tot}
                = \sum_{i \in \Gamma_A} (S^z_i + S^z_{i + \delta})
                = \sum_{i \in \Gamma_A} (\abos*{i}\abos{i} - \bbos*{i + \delta}\bbos{i + \delta}).
			\ee
      From this conservation we can infer
      \begin{align}
                \cexpval{ \abos{i} \abos{i} } = \cexpval{ \bbos*{i+\delta} \bbos*{i+\delta} } 
								= \cexpval{ \abos{i} \bbos*{i+\delta} } = \cexpval{ \abos*{i} \bbos{i+\delta} } = 0.
      \end{align}
      In total, the mean-field decoupled Hamiltonian reads
      \begin{align}
        \mathcal{H} &= J \sum_{i \in \Gamma_A, \delta} 
								\Big[  -S^2 + 2Sn + 2S\lambda \Delta - n^2 
								 - \Delta^2 - 2\lambda n \Delta \nonumber\\ 
        &\qquad+ \left( S - n -\lambda \Delta \right) \big( \norord{ \abos*{i} \abos{i} }
								 + \norord{\bbos*{i+\delta} \bbos{i+\delta}} \big)  \nonumber \\
          &\qquad +\left(\lambda S -\lambda n-\Delta\right) \big( \norord{\abos{i}\bbos{i+\delta}}
								 + \norord{\abos*{i}\bbos*{i+\delta}} \big) \nonumber \\
          &\qquad- \norord{ \abos*{i} \abos{i} \bbos*{i+\delta}  \bbos{i+\delta} }
						\\
          &\qquad  -\frac{\lambda}{2} \norord{ \abos*{i} \abos{i}  \abos{i} \bbos{i+\delta} }
								 -\frac{\lambda}{2} \norord{ \abos*{i} \bbos*{i+\delta}  
								\bbos*{i+\delta} \bbos{i+\delta} } \Big] \nonumber.
      \end{align}
      The first line is the zero-magnon part, the second and third line the one-magnon part and the last two lines comprise the two-magnon terms. We stress that no approximation is performed.
      Next, we apply a Fourier transformation yielding
      \begin{align}
                \mathcal{H} &= J  \Big[  N \left( -S^2 + 2Sn + 
								2S\lambda \Delta - n^2 - \Delta^2 - 2\lambda n \Delta \right)   
								\nonumber  \\ 
                &+ \sum_{\vect{k}\in \text{MBZ}} A_{\vect{k}} \big( \norord{ \abos*{\vect{k}} 
														\abos{\vect{k}} } 
								+ \norord{\bbos*{\vect{k}} \bbos{\vect{k}}} \big) 
														\nonumber \\
                &+ \sum_{\vect{k}\in\text{MBZ}} B_{\vect{k}} \big( \norord{ \abos{\vect{k}} 
														\bbos{-\vect{k}}}
														+ \norord{\abos*{\vect{k}}\bbos*{-\vect{k}}} \big)  
														\\
                &- \frac{Z}{N} \sum_{\vect{1},\vect{2},\vect{3},\vect{4}\in\text{MBZ}}
								     \delta(\vect{1}\!-\!\vect{2}\!+\!\vect{3}\!-\!\vect{4}) 
										 \gamma(\vect{3}\!-\!\vect{4}) 
										 \norord{ \abos*{\vect{1}} \abos{\vect{2}} \bbos*{\vect{3}} \bbos{\vect{4}}} 														\nonumber\\
                & -\frac{\lambda Z}{2 N} \sum_{\vect{1},\vect{2},\vect{3},\vect{4}\in\text{MBZ}} 
														\delta(\vect{1}\!-\!\vect{2}\!-\!\vect{3}\!-\!\vect{4})
														\gamma(\vect{4})\norord{ \abos*{\vect{1}} \abos{\vect{2}} 
														\abos{\vect{3}} \bbos{\vect{4}} } \nonumber \\
                            &  -\frac{\lambda Z}{2 N} \!\!\! 
														\sum_{\vect{1},\vect{2},\vect{3},\vect{4}\in\text{MBZ}}
														\!\!\!\!\!\!\!\!\!
														\delta(\vect{1}\!+\!\vect{2}\!+\!\vect{3}\!-\!\vect{4}) 
														\gamma(\vect{4}\!-\!\vect{2}\!-\!\vect{3})
														\norord{ \abos*{\vect{1}} \bbos*{\vect{2}} \bbos*{\vect{3}} 
														\bbos{\vect{4}}}\!\Big], \nonumber
														\label{a:eq:quad}
    \end{align}
    where we use the shorthand $\vect{j}$ for $\vect{k}_j$.
    The Kronecker $\delta$ symbols hold modulo reciprocal lattice vectors.
    If a term describes an Umklapp process, i.e., a reciprocal lattice vector $\vect{G}$ is needed to fulfill momentum conservation, the Kronecker $\delta$ takes the value $\gamma(\vect{G}) = \pm 1$.
    In the above equation, we used the definitions
    \bes
    \begin{align}
        \gamma(\vect{k}) &\coloneqq \frac{1}{Z} \sum_{\delta} {\text{e}}^{-\text{i}\delta \vect{k}} 
                            = \frac{2}{Z} \left[ \cos(k_x) + \cos(k_y) \right]  \\
            A_{\vect{k}} &\coloneqq Z \left( S - n -\lambda \Delta \right) \\
            B_{\vect{k}} &\coloneqq  Z \gamma(\vect{k}) \left(\lambda S -\lambda n-\Delta \right),
    \end{align}
        \ees
    where we introduced the coordination number $Z=4$.
    Next, we diagonalize the bilinear part in Eq.\ \eqref{a:eq:quad} by the Bogoliubov transformation
    \bes
    \begin{align}
                \abos*{\vect{k}} &= l_{\vect{k}} \alp*{\vect{k}} + m_{\vect{k}} \bet{-\vect{k}} \\
                \bbos*{\vect{k}} &= m_{\vect{k}} \alp{\vect{-k}} + l_{\vect{k}} \bet*{\vect{k}}
      \end{align}
	\ees
    with
    \be
        l_{\vect{k}}^2 -  m_{\vect{k}} ^2 = 1.
    \ee
    Explicitly, we parametrize as in Refs.\ \cite{Powalski_2015, Powalski_2018, Uhrig_2013}
	\bes
        \begin{align}
             \mu_{\vect{k}} &\coloneqq \sqrt{1 - \left( \frac{B_{\vect{k}}}{A_{\vect{k}}} \right)^2 } \\
             l_{\vect{k}} &\coloneqq \sqrt{ \frac{1- \mu_{\vect{k}}}{2  \mu_{\vect{k}}} }
        \end{align}
        \begin{align}
             m_{\vect{k}} &\coloneqq - \sign(\gamma(\vect{k})) 
                             \sqrt{ \frac{1 + \mu_{\vect{k}}}{2  \mu_{\vect{k}}} } 
                             = - \sign(\gamma(\vect{k})) x_{\vect{k}} l_{\vect{k}} \\
             x_{\vect{k}} &\coloneqq \sqrt{ \frac{1+ \mu_{\vect{k}}}{1- \mu_{\vect{k} } } }.             
         \end{align}
	\ees
	With these steps we rewrite the  Hamiltonian \eqref{eq:startHamil} in the form
    \begin{align}
        \mathcal{H} = E_0 + \sum_{\vect{k}} \disp_0(\vect{k}) 
                        \big(  \norord{ \alp*{\vect{k}}\alp{\vect{k}}} 
                        + \norord{\bet*{\vect{k}}\bet{\vect{k}}}  \big) + \Gamma_0 + \mathcal{V}_0,
      \end{align}
      with
			\bes
    \begin{align}
        E_0 &= J N \left( -S^2 + 2Sn + 2S\lambda \Delta - n^2 - \Delta^2 - 
                            2\lambda n \Delta \right) \\
        \omega_0(\vect{k}) &= A_{\vect{k}} \mu_{\vect{k}} \\
        \Gamma_0 &= \sum_{\vect{k}\in \text{MBZ}} \left( 2 A_{\vect{k}} l_{\vect{k}} m_{\vect{k}} 
                    + B_{\vect{k}}( m_{\vect{k}}^2 + l_{\vect{k}}^2 ) \right) 
                    \nonumber\\
                        &\qquad\qquad \times 
        \big( \norord{ \alp{\vect{k}}\bet{-\vect{k}}} + 
                \norord{\alp*{\vect{k}}\bet*{-\vect{k}}} \big).
    \end{align}
    \ees
    The term $\mathcal{V}_0$ comprises all the quadrilinear two-magnon terms.
    We do not provide here explicitly because is already published, for instance in Ref.\ \cite{Uhrig_2013}, except that we denote the normal-ordering of the quadrilinear terms by $\norord{...}$, and add a factor $\lambda$ in front of $\gamma$ with one or a sum of three momenta as arguments.
    For example the term $x_1 \gamma(1)$ becomes $ \lambda x_1 \gamma(1)$, the term $x_1 x_3 x_4 \gamma(1-3-4)$ becomes $ \lambda x_1 x_3 x_4 \gamma(1-3-4)$, while $x_1 x_3 \gamma(1-3)$ stays $x_1 x_3 \gamma(1-3)$. 
    This can be directly traced back to the addtional $\lambda$ factors in the quartic part of Eq. \ref{a:eq:quad} compared to the Heiseberg model.

    In order to retrieve the values of $n$ and $\Delta$, we exploit translational invariance and finally perform the Fourier and Bogoliubov transformation on the right hand sides of Eqs.\ \eqref{eq:n_Delta}, evaluate the vacuum expectation value for the Bogoliubov bosons 
    $\alp{\vect{k}}$ and $\bet{\vect{k}}$ to obtain the self-consistency conditions
    \bes
    \begin{align}
        n &= \frac{1}{N} \sum_{\vect{k}} \cexpval{ \abos*{\vect{k}} \abos{\vect{k}} } 
                            =\frac{1}{N} \sum_{\vect{k}} l_{\vect{k}}^2 \\
        \Delta &= \frac{1}{N} \sum_{\vect{k}}\gamma(\vect{k}) 
                        \cexpval{ \abos*{\vect{k}} \bbos{\vect{k}} } 
                    = \frac{1}{N} \sum_{\vect{k}} \gamma(\vect{k}) l_{\vect{k}} m_{\vect{k}}.
    \end{align}
        \ees
    They are solved numerically by iteration to convergence for $\mbz$ mesh with $L=30001$ sites with a tolerance $10^{-13}$.
    \end{appendix}

\end{document}